\documentclass[conference,letterpaper]{IEEEtran}

\usepackage{graphicx}
\usepackage[cmex10]{amsmath}
\usepackage{amsfonts}
\usepackage{amsthm}
\usepackage{bbm}

\usepackage[ruled,vlined]{algorithm2e}
\usepackage{mathtools}

\def\x{{\mathbf x}}
\def\X{{\mathbf X}}
\def\y{{\mathbf y}}
\def\z{{\mathbf z}}
\def\w{{\mathbf w}}

\def\t{{\mathbf t}}
\def\u{{\mathbf u}}

\newcommand{\B}[1]{\ensuremath{\mathbf{#1}}} 
\newcommand{\norm}[1]{\ensuremath{\|#1\|}} 
\def\R{{\mathbb R}}
\def\Z{{\mathbb Z}}

\def\cX{{\mathcal X}}

\usepackage[utf8]{inputenc} 
\usepackage[T1]{fontenc}
\usepackage{url}
\usepackage{ifthen}
\usepackage{cite}

\theoremstyle{definition}
\newtheorem{definition}{Definition}

\theoremstyle{plain}
\newtheorem{theorem}{Theorem}

\newtheorem{corollary}{Corollary}
\newtheorem{prop}{Proposition}

\IEEEoverridecommandlockouts

\begin{document}
\title{An MCMC Method to Sample from Lattice Distributions}

\author{
  \IEEEauthorblockN{Anand Jerry George}
  \IEEEauthorblockA{Department of Electrical Engineering\\
  Indian Institute of Science, Bengaluru\\
                    Email: anandgeorge@iisc.ac.in}
  \and 
  
  \IEEEauthorblockN{Navin Kashyap}
  \IEEEauthorblockA{Department of Electrical Communication Engineering\\
  Indian Institute of Science, Bengaluru\\
  Email: nkashyap@iisc.ac.in}
  \thanks{The work of N. Kashyap was supported in part by a Swarnajayanti Fellowship awarded by the Dept. of Science and Technology (DST), Govt. of India.}
}

\maketitle

\begin{abstract}
We introduce a Markov Chain Monte Carlo (MCMC) algorithm to generate samples from probability distributions supported on a $d$-dimensional lattice $\Lambda = \B{B}\Z^d$, where $\B{B}$ is a full-rank matrix. Specifically, we consider lattice distributions $P_\Lambda$ in which the probability at a lattice point is proportional to a given probability density function, $f$, evaluated at that point. To generate samples from $P_\Lambda$, it suffices to draw samples from a pull-back measure $P_{\Z^d}$ defined on the integer lattice. The probability of an integer lattice point under $P_{\Z^d}$ is proportional to the density function $\pi = |\det(\B{B})|f\circ \B{B}$. The algorithm we present in this paper for sampling from $P_{\Z^d}$ is based on the Metropolis-Hastings framework. In particular, we use $\pi$ as the proposal distribution and calculate the Metropolis-Hastings acceptance ratio for a well-chosen target distribution. We can use any method, denoted by \textbf{ALG}, that ideally draws samples from the probability density $\pi$, to generate a proposed state. The target distribution is a piecewise sigmoidal distribution, chosen such that the coordinate-wise rounding of a sample drawn from the target distribution gives a sample from $P_{\Z^d}$. When \textbf{ALG} is ideal, we show that our algorithm is uniformly ergodic if $-\log(\pi)$ satisfies a gradient Lipschitz condition.
\end{abstract}
\section{Introduction}
\label{sec:intro}
Drawing samples from probability distributions is a ubiquitous problem in statistics and machine learning. In this paper, we look into the problem of sampling from probability distributions supported on a lattice (referred to as a \textit{lattice distribution}). Specifically, we consider lattice distributions in which the probability at a lattice point is proportional to a given probability density function evaluated at that point.  Such distributions have found applications in cryptography, lattice coding, and secure communication\cite{Gentry_Peikert},\cite{AWGN_LING},\cite{Vatedka}. An example of a lattice distribution that has received much interest from researchers is the lattice Gaussian. A lattice Gaussian distribution is a probability distribution defined on a lattice $\Lambda$, such that for each $\x\in\Lambda$, the probability of $\x$ is proportional to a Gaussian density function evaluated at $\x$. Lattice Gaussian sampling, also known as discrete Gaussian sampling (DGS), is closely related to shortest vector problem (SVP) and closest vector problem (CVP), which are computationally hard lattice problems\cite{dgs}. Lattice Gaussian sampling is also employed for decoding and signal detection in Multiple Input Multiple Output (MIMO) systems\cite{Decoding_Ling}. There are some known methods to generate samples from lattice Gaussian distributions\cite{Gentry_Peikert},\cite{wang_ling},\cite{learnableMCMC}, but few methods are available for drawing samples from arbitrary lattice distributions\cite{MTech_rep_Anaswara}. We try to address this problem. The motivation to go beyond lattice Gaussian is due to \cite{Vatedka}, where samples from a particular fat-tailed lattice distribution are used to achieve information-theoretically perfect security. 

Since sampling methods for generic lattice distributions are not known prior to this work, we compare our algorithm with other available lattice Gaussian samplers. Algorithms currently available for sampling from $d$-dimensional lattice Gaussians require a sub-routine to generate samples from a 1-dimensional lattice Gaussian with arbitrary variance parameter. These schemes update each coordinate sequentially and therefore, cannot take advantage of the currently available optimized linear algebra libraries. Also, techniques used to sample from 1-dimensional lattice Gaussians are either computationally inefficient or require a pre-computed table\cite{1D_DGS_Follath}. This paper proposes a simple and provable algorithm that samples directly from $d$-dimensional lattice distributions, including lattice Gaussians.

 Now we give a brief introduction to our algorithm. Let $\Lambda$ be a $d$-dimensional lattice generated by a full-rank matrix $\B{B}$. As mentioned earlier, we consider lattice distributions $P_\Lambda$ in which the probability at a lattice point is proportional to the value of a given probability density function $f$ at that point. By a simple reduction, it becomes evident that, to generate samples from $P_\Lambda$, it suffices to draw samples from a probability distribution $P_{\Z^d}$ defined on the integer lattice. The probability of an integer lattice point under $P_{\Z^d}$ is proportional to the probability density function
 \begin{equation}\label{pi_intro}
      \pi(\x) = |\det(\B{B})|f(\B{B}\x) \quad \text{for all } \x\in\R^d.
 \end{equation}
 This paper explores the possibility of using Markov chain Monte Carlo (MCMC) methods to draw samples from $P_{\Z^d}$. MCMC is a well-known paradigm in statistics to sample from the desired probability distribution by establishing a Markov chain whose stationary distribution is the same as the desired distribution. In particular, we use the well-known Metropolis-Hastings algorithm to establish a Markov chain with the desired stationary distribution (referred to as the \textit{target distribution}). In a Metropolis-Hastings algorithm, a candidate for the next state of the Markov chain is drawn from a particular \textit{proposal distribution} and is then accepted as the next state with a certain \textit{acceptance probability}. In contrast to the other instances in literature where Metropolis-Hastings algorithms are used for lattice Gaussian sampling\cite{wang_ling},\cite{learnableMCMC}, we use a proposal distribution and target distribution that have probability density functions. In particular, we use $\pi$ given in (\ref{pi_intro}) as the proposal distribution. We can use any off-the-shelf method, denoted by \textbf{ALG}, that ideally draws samples from the probability density $\pi$, to generate a proposed state. In fact, this enables us to sample from lattice distributions beyond lattice Gaussians, i.e., when $\pi$ is not a Gaussian density. We use a piecewise sigmoidal probability density as the target distribution, a sample from which, after coordinate-wise rounding, gives a sample from $P_{\Z^d}$. With this choice of the target distribution, we show that our algorithm has exponentially fast convergence (\textit{uniform ergodicity}) if the lattice distribution is such that $-\log(\pi)$ satisfies a gradient Lipschitz condition. We also derive a bound on the rate at which our algorithm converges to the target distribution when \textbf{ALG} deviates from its ideality.
\subsection{Organization of the Paper}
The remainder of the paper is organized as follows. In Section~\ref{preliminaries}, we formalize the problem statement and recall some basics of MCMC and its convergence. In Section~\ref{IMHR_section}, we describe our algorithm and its convergence analysis. Proposition~\ref{uniErg} in Section~\ref{IMHR_section} shows that our algorithm is uniformly ergodic if $-\log(\pi)$ satisfies a gradient Lipschitz condition. In Proposition~\ref{inexactProp}, we derive a bound on the rate at which our algorithm converges to the target distribution when \textbf{ALG} is non-ideal. In Section~\ref{SimRes}, we present our algorithm's simulation results for three different target distributions: isotropic Gaussian distribution on $\Z^d$,  isotropic Gaussian distribution on the Leech lattice, and ``Perfect Security distribution'' on $\Z^d$. Section~\ref{discussion} discusses the performance of our algorithm and compares it with Klein's algorithm\cite{klein},\cite{Gentry_Peikert} for lattice Gaussian sampling.

\section{Preliminaries}\label{preliminaries}
\subsection{Notations}
We denote the state space of a Markov chain by $\mathcal{X}$. The operator $\wedge$ operates on two numbers to output the minimum of them. Nearest integer point to a vector $\x\in\R^d$ obtained by coordinate-wise rounding is denoted by $[\x]$. We use $U[0,1]$ to denote the uniform probability distribution on the interval $[0,1]$.  We denote the Borel-sigma algebra on $\R^d$ by $\mathcal{B}(\R^d)$. For two probability measures $\mu$ and $\nu$ defined on the same probability space, we use the notation $\mu\ll\nu$ to indicate that $\mu$ is absolutely continuous with respect to $\nu$.

\subsection{Lattice Distributions}
We now formally define a lattice and probability distributions defined on a lattice. Let $\B{B} \in \R^{d\times d}$ be a full-rank matrix. The $d$-dimensional lattice $\Lambda$  generated by $\B{B}$ is defined as 
\begin{equation*}
    \Lambda \coloneqq \{\B{Bz}:\z\in \Z^d\}.
\end{equation*}
Any probability distribution defined with $\Lambda$ as the support is known as a lattice distribution. In this paper, we look at a specific class of lattice distributions in which the probability distribution is induced by a density function on $\R^d$. That is, the probability of a lattice point is equal to the density function evaluated at that point with appropriate normalization. This paper mainly considers density functions having the following form
\begin{equation*}
    f(\x) = \frac{e^{-\psi(\x)}}{Z_{\psi}} \quad \text{for all } \x\in \R^d,
\end{equation*}
where $\psi(\x)$ is known as a \textit{potential function}, and $Z_{\psi} = \int_{\R^d} e^{-\psi(\x)} d\x$ is a normalization constant. However, in practice, the algorithm that we develop works for any lattice distribution induced by a density function. Let $P_{\Lambda}(\x)$ for $\x\in\Lambda$ be a lattice distribution induced by the above $f(\x)$, i.e.,
\begin{equation*}
    P_{\Lambda}(\x) = \frac{e^{-\psi(\x)}}{Z} \quad \text{for all } \x\in \Lambda,
\end{equation*}
where
\begin{equation*}
    Z = \sum_{\x\in\Lambda}e^{-\psi(\x)}.
\end{equation*}
Let $\B{B}$ denote a generator matrix of the lattice $\Lambda$. Then, for generating a sample from the probability distribution $P_\Lambda$, it suffices to sample from $P_{\Z^d}(\z) = P_\Lambda(\B{Bz})$, where $\z\in\Z^d$, and then obtain $\x$ as $\x = \B{Bz}$. So, our problem reduces to one of sampling from the following probability distribution over $\Z^d$:
\begin{equation*}
    P_{\Z^d}(\z) = \frac{e^{-\psi(\B{Bz})}}{Z} \quad \text{for all } \z\in \Z^d.
\end{equation*}
Let $\varphi(\x) \coloneqq \psi(\B{Bx})$ for all $\x\in\R^d$, so that
\begin{equation}\label{PZd}
      P_{\Z^d}(\z) = \frac{e^{-\varphi(\z)}}{Z} \quad \text{for all } \z\in \Z^d.  
\end{equation}
Also, let us define the probability density $\pi$ as:  
\begin{equation}\label{pi}
    \pi(\x) \coloneqq \frac{e^{-\varphi(\B{x})}}{K} \quad \text{for all } \x\in \R^d,
\end{equation}
where
\begin{equation*}
    K = \int_{\R^d}e^{-\varphi(\x)}d\x.
\end{equation*}
Note that $\pi$ is related to $f$ as given in (\ref{pi_intro}). We have now reduced the problem of sampling from a probability distribution defined on an arbitrary lattice $\Lambda$ to sampling from a probability distribution defined on $\Z^d$. In the rest of this paper, we try to develop an algorithm for generating samples from the lattice distribution $P_{\Z^d}$ induced by the probability density $\pi$. 

\subsubsection{Lattice Gaussian Distribution}
A lattice distribution that has received significant attention is the lattice Gaussian distribution. A lattice Gaussian distribution defined on a lattice $\Lambda$ is given by
\begin{equation}\label{latticeGaussian}
    D_{\Lambda,\sigma,\B{c}}(\x) = \frac{e^{-\frac{\|\x-\B{c}\|^2}{2\sigma^2}}}{\sum_{\y\in \Lambda}e^{-\frac{\|\y-\B{c}\|^2}{2\sigma^2}}} \quad \text{for all } \x \in \Lambda,
\end{equation}
where $\B{c}\in\R^d$ is the mean vector and $\sigma$ is the variance parameter. Lattice Gaussian distributions have important practical applications, particularly in cryptography\cite{Gentry_Peikert}.

\subsection{The Metropolis-Hastings Algorithm}
As stated earlier, we take the MCMC route to generate samples from the desired lattice distribution. MCMC is a class of sampling algorithms in which a Markov chain is set up whose stationary distribution is the same as the desired probability distribution (also called the \textit{target distribution}). The idea is to simulate this chain for a certain number of steps to draw samples approximately from the desired probability distribution.

Given a Markov chain, it is straightforward to find its stationary distribution. However, it is not apparent how to find a Markov chain with the desired stationary distribution. The Metropolis-Hastings algorithm provides a recipe for establishing a Markov chain with the desired stationary distribution. Let $\bar{\pi}$ denote the probability distribution from which we want to draw samples. Then, the Metropolis-Hastings algorithm consists of two steps in generating the next state of the Markov chain:
\begin{itemize}
    \item Let $\x$ be the current state. Generate a proposed state $\y$ from some probability distribution $q(\x,\cdot)$ (referred to as the \textit{proposal distribution}).
    \item Accept the proposed state as the next state of Markov chain with probability $\alpha(\x,\y)$ given by
    \begin{equation*}
        \alpha(\x,\y) = 1\wedge\frac{\bar{\pi}(\y)q(\y,\x)}{\bar{\pi}(\x)q(\x,\y)}.
    \end{equation*}
\end{itemize}
If the proposed state is independent of the current state, we call this the \textit{Independent Metropolis-Hastings} algorithm. The acceptance ratio is then given by
    \begin{equation*}
        \alpha(\x,\y) = 1\wedge\frac{\bar{\pi}(\y)q(\x)}{\bar{\pi}(\x)q(\y)}.
    \end{equation*}
From now on, we refer to the Markov chain associated with an Independent Metropolis-Hastings algorithm by MH Markov chain.

\subsection{Distance between probability distributions}
For assessing the goodness of any sampling algorithm, it is essential to have a metric defined on the space of probability distributions. The metric we use in our analysis is the Total Variation Distance (TVD). For two distributions $\mu$ and $\nu$ defined on $(\R^d,\mathcal{B}(\R^d))$, we use $\norm{\mu-\nu}_{TV}$ to denote their TVD given by
 \begin{equation*}
     \norm{\mu-\nu}_{TV} = \sup_{A\in\mathcal{B}(\R^d)}|\mu(A)-\nu(A)|.
 \end{equation*}
If $\lambda$ is a probability measure such that $\mu\ll\lambda$ and $\nu\ll\lambda$, then an alternate expression for TVD is given by
 \begin{equation}\label{TVD_radonNik}
     \norm{\mu-\nu}_{TV} = \frac{1}{2}\int_{\R^d}\left|\frac{d\mu}{d\lambda}(\x)-\frac{d\nu}{d\lambda}(\x)\right|\lambda(d\x),
 \end{equation}
 where $\frac{d\mu}{d\lambda}$ and $\frac{d\nu}{d\lambda}$ are the Radon-Nikodym derivatives of $\mu$ and $\nu$ with respect to $\lambda$ (see Lemma 2.1 in \cite{nonParamTsybakov}).

\subsection{Convergence to stationarity}
In this section, we give some definitions useful in evaluating the convergence of a Markov chain to its stationary distribution. We refer the reader to \cite{MCSS_Meyn_Tweedie},\cite{Roberts_2004} for a comprehensive review of these topics.


\begin{definition}
 A Markov chain with transition kernel $P$ and stationary distribution $\bar{\pi}$ is \textit{uniformly ergodic} if there exists $0<\delta<1$ and $M<\infty$ such that for all $\x\in\mathcal{X}$,
 \begin{equation*}
     \norm{P^t(\x,\cdot)-\bar{\pi}(\cdot)}_{TV} \le M(1-\delta)^t.
 \end{equation*}
\end{definition}
\begin{theorem}\label{compState_uniform}
(Theorem 8 in \cite{Roberts_2004}). Let $P$ be the transition kernel of a Markov chain and $\bar{\pi}$ be its stationary distribution. Suppose there exists a $\delta>0$ and a probability measure $\nu$ such that, for all measurable $B\subseteq\mathcal{X}$, 
\begin{equation*}
    P(\x,B)\ge \delta \nu(B) \quad \text{for all } \x\in \mathcal{X}. 
\end{equation*}
Then the Markov chain with transition kernel $P$ is uniformly ergodic and satisfies the following inequality:
\begin{equation*}
         \norm{P^t(\x,\cdot)-\bar{\pi}(\cdot)}_{TV} \le (1-\delta)^t \quad \text{for all } \x\in\mathcal{X}.
\end{equation*}
\end{theorem}
\begin{definition}
 A Markov chain with transition kernel $P$ and stationary distribution $\bar{\pi}$ is \textit{geometrically ergodic} if there exists $0 < \delta < 1$ such that, for all $\x \in \cX$,
 \begin{equation*}
     \norm{P^t(\x,\cdot)-\bar{\pi}(\cdot)}_{TV} \le M(\x)(1-\delta)^t,
 \end{equation*}
with $M(\x) < \infty$.
\end{definition}

\begin{definition}
For a small $\epsilon>0$, and initial state $\x$, a \textit{mixing time} $t_{\text{mix}}(\epsilon;\x)$ is defined as 
\begin{equation*}
    t_{\text{mix}}(\epsilon;\x) = \inf\{t : \norm{P^t(\x,\cdot)-\bar{\pi}(\cdot)}_{TV} < \epsilon\}.
\end{equation*}
\end{definition}

\section{Independent Metropolis-Hastings with Rounding (IMHR)}\label{IMHR_section}
In this section, we introduce an Independent Metropolis-Hastings algorithm for sampling from lattice distribution $P_{\Z^d}$ defined in (\ref{PZd}). In this algorithm, we suppose that it is possible to generate samples from the probability density $\pi$ defined in (\ref{pi}). Any state-of-the-art MCMC algorithm such as Hamiltonian Monte Carlo (HMC)\cite{neal2012mcmc}, or Metropolis adjusted Langevin algorithm (MALA)\cite{Langevin_roberts1996}, can be used for this purpose. The idea is to use $\pi$ as the proposal distribution in the Independent Metropolis-Hastings algorithm. For such a method to be effective in sampling from $P_{\Z^d}$, we need a target distribution $\bar{\pi}$ with the following properties:
\begin{itemize}
    \item Using a random variable with probability distribution $\bar{\pi}$, we should be able to efficiently derive a random variable with distribution $P_{\Z^d}$.
    \item The probability distribution $\bar{\pi}$ should be statistically close to $\pi$. This will reduce the possibility of rejecting a proposal in the Independent Metropolis-Hastings algorithm, thereby improving its convergence speed to the stationary distribution. 
\end{itemize}

A naive approach would be to choose $\bar{\pi}$ as a piece-wise constant density. That is, $\bar{\pi}(\x)$ is equal to $\pi([\x])$ with appropriate normalization for all $\x\in\R^d$. It is easy to see that the rounding operation on a sample generated from $\bar{\pi}$ gives a sample from $P_{\Z^d}$. The drawback of such an approach is that the Markov chain thus generated need not be uniformly ergodic, even for lattice Gaussians (see Appendix~\ref{PWConstApprox}). This motivates us to find a $\bar{\pi}$ that is a better approximation to $\pi$. 

\begin{figure}[htb]
\centering
\includegraphics[scale=0.47]{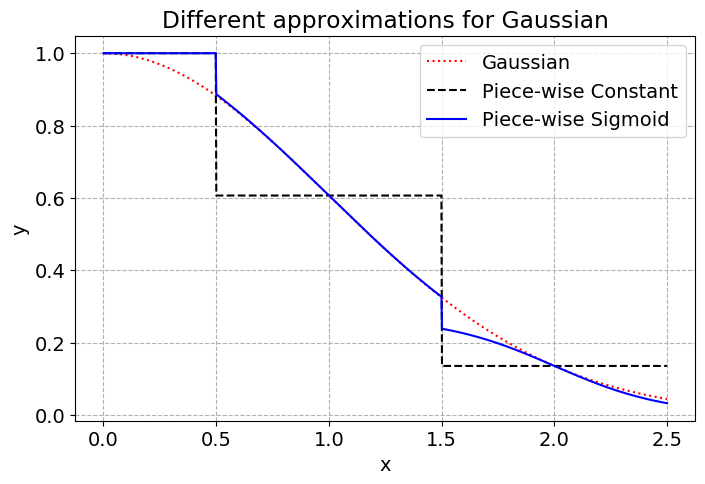}
\caption{Different approximations for Gaussian density}
\label{sigmoid}
\end{figure}
We define a new probability distribution $\bar{\pi}$ which will be called the \textit{target distribution} henceforth, as follows:
\begin{equation}\label{piBar}
    \bar{\pi}(\x) =\frac{1}{Z} \frac{2e^{-\varphi(\bar{\x})}}{1+e^{2(\x-\bar{\x})^T\nabla\varphi(\bar{\x})}}\quad \text{for all } \x\in\R^d,
\end{equation}
where $Z$ and $\varphi$ are the same entities which appear in (\ref{PZd}), and $\bar{\x}$ is the nearest integer point to $\x$ which is obtained by coordinate-wise rounding. We should visualize $\bar{\pi}$ as a probability density function obtained by approximating $\pi$ using a sigmoid function within each unit hypercube in $\R^d$ and then normalizing. The sigmoid function is chosen such that, at the center of any unit hypercube, its value and gradient are proportional to the value and gradient of $\pi$. This is illustrated in Figure \ref{sigmoid} where $\pi$ is a Gaussian density function. Note that the functions plotted in Figure~\ref{sigmoid} are unnormalized. We can obtain a sample from $P_{\Z^d}$ by rounding the sample generated from $\bar{\pi}$ to its nearest integer point. To see this, let $\X$ be a random variable with probability density $\bar{\pi}$. Let $\B{Z} = [\X]$ and let $S$ denote the unit hypercube centered at the origin, i.e., $S = [-\frac{1}{2},\frac{1}{2}]^d$. Then,
\begin{equation*}
\begin{split}
   \mathbb{P}&(\B{Z}=\z) = \mathbb{P}(\X\in\z+S) \\
   &= \int_{S}\bar{\pi}(\z+\u)d\u \\
   &\stackrel{(a)}= \frac{1}{2}\int_{S}\left(\bar{\pi}(\z+\u)+\bar{\pi}(\z-\u)\right)d\u \\
   &= \frac{1}{Z}e^{-\varphi(\z)}\int_{S} \left(\frac{1}{1+e^{2\u^T\nabla\varphi(\z)}}+ \frac{1}{1+e^{-2\u^T\nabla\varphi(\z)}}\right)d\u \\
  &= \frac{1}{Z}e^{-\varphi(\z)}\int_{S} \left(\frac{1}{1+e^{2\u^T\nabla\varphi(\z)}}+ \frac{e^{2\u^T\nabla\varphi(\z)}}{1+e^{2\u^T\nabla\varphi(\z)}}\right)d\u \\
   &= \frac{1}{Z}e^{-\varphi(\z)}, \\
\end{split}
\end{equation*}
where $(a)$ is due to the symmetry of $S$. Therefore, to generate samples from $P_{\Z^d}$, it suffices to draw samples from $\bar{\pi}$ and then do coordinate-wise rounding. 

 Summarizing, IMHR is an Independent Metropolis-Hastings algorithm with $\pi$ defined in (\ref{pi}) as the proposal distribution and $\bar{\pi}$ defined in (\ref{piBar}) as the target distribution. The steps of IMHR are as described in Algorithm \ref{IMHR}.\\

\begin{algorithm}\label{IMHR}
\SetAlgoLined
\KwIn{ $\X_0,\pi,\varphi,\nabla\varphi$}
\KwOut{Sample from a distribution statistically close to $P_{Z^d}$}
 \For{$t=1,2,\ldots$}{
  Let $\x$ be the state of $\X_{t-1}$\;
  Generate $\y$ from the probability distribution $\pi$\;
  Round $\x$ to its nearest point in $\Z^d$ to get $\bar{\x}$\;
  Round $\y$ to its nearest point in $\Z^d$ to get $\bar{\y}$\;
  $\bar{\pi}(\x) = \frac{2\exp(-\varphi(\bar{\x}))}{1+\exp(2(\x-\bar{\x})^T\nabla\varphi(\bar{\x}))}$\;
  $\bar{\pi}(\y) = \frac{2\exp(-\varphi(\bar{\y}))}{1+\exp(2(\y-\bar{\y})^T\nabla\varphi(\bar{\y}))}$\;
  Calculate acceptance ratio $\alpha(\x,\y) = 1\wedge\frac{\bar{\pi}(\y)\pi(\x)}{\bar{\pi}(\x)\pi(\y)}$\;
  Generate a sample $u$ from $U[0,1]$\;
  \eIf{$u\le \alpha(\x,\y)$}{
   let $\X_t=\y$\;
   }{
   $\X_t=\x$\;
  }
  \If{$t>t_{\text{mix}}(\epsilon;\X_0)$}{
  Round $\X_t$ to its nearest point in $\Z^d$ to get $\bar{\X}_t$\;
  Output $\bar{\X}_t$;
  }
 }
 \caption{IMHR Algorithm}
\end{algorithm}

\subsection{Convergence analysis of Algorithm \ref{IMHR}}
In this section, we analyze the convergence speed of Algorithm~\ref{IMHR} to its stationary distribution. Algorithm~\ref{IMHR} requires a sub-routine, denoted by \textbf{ALG} henceforth, which is ideally capable of drawing samples from $\pi$. The following analysis assumes that we have such a sub-routine available. Error due to non-availability of such an ideal sub-routine will be analyzed in the next section.

First, we state a well known theorem which is true in general for an Independent Metropolis-Hastings algorithm. 

\begin{theorem}\label{qbipiThm}
(Theorem 2.1 in \cite{mengersen1996}) An Independent Metropolis-Hastings algorithm is uniformly ergodic if there exist $\delta>0$ such that 
\begin{equation}\label{qbipi}
\frac{\pi(\x)}{\bar{\pi}(\x)}\ge \delta \quad \text{for all }\x\in\mathcal{X},
\end{equation}
where $\pi$ is the density from which proposed state is generated, and $\bar{\pi}$ is the target density.
That is, the transition kernel $\bar{P}$ of the MH Markov chain satisfies the following:
\begin{equation*}
     \|\bar{P}^k(\x,\cdot)-\bar{\pi}(\cdot)\|_{TV} \le (1-\delta)^k \quad \text{for all }\x\in\mathcal{X}.
\end{equation*}
\end{theorem}
Next, we define a widely used smoothness property of functions called $L$-smoothness.
\begin{definition}
A function $f:\R^d\rightarrow\R$ is called \textit{$L$-smooth} if the gradient of $f$ is Lipschitz continuous with parameter $L$. That is, $f$ should satisfy the following property
\begin{equation*}
    \norm{\nabla f(\x)-\nabla f(\y)} \le L\norm{\x-\y}, \quad \text{for all } \x,\y\in\R^d.
\end{equation*}
\end{definition}
The following is a well known fact about $L$-smooth functions (see Lemma~5 in \cite{Dwivedi_MALA}). If $f$ is $L$-smooth, then for all $\x,\y\in\R^d$,
\begin{equation}\label{Lsmooth}
    f(\y)\le f(\x)+(\y-\x)^T\nabla f(\x)+\frac{L}{2}\norm{\x-\y}^2.
\end{equation}
Now we will give conditions on the probability density $\pi$ that guarantees uniform ergodicity for Algorithm~\ref{IMHR}.

\begin{prop}\label{uniErg}
Let $\pi$ and $\bar{\pi}$ be as defined in (\ref{pi}) and (\ref{piBar}) respectively. Let $\bar{\x}$ denote the nearest integer point to $\x$. If $\varphi$ is an $L$-smooth function, then for all $\x\in\R^d$ we have,
\begin{equation}\label{pibpib}
   \frac{\pi(\x)}{\bar{\pi}(\x)} = \frac{Z}{K}\frac{e^{-\varphi(\x)}(1+e^{2(\x-\bar{\x})^T\nabla\varphi(\bar{\x})})}{2e^{-\varphi(\bar{\x})}} \ge \frac{Z}{K}e^{-\frac{dL}{8}} > 0.
\end{equation}
Therefore, by Theorem~\ref{qbipiThm}, Algorithm~\ref{IMHR} is uniformly ergodic for such a $\pi$.
\end{prop}
\begin{IEEEproof}
 Let $\y = \x-\bar{\x}$. Then,
    \begin{equation*}
\begin{split}
   \frac{\pi(\x)}{\bar{\pi}(\x)} &= \frac{Z}{K}\frac{e^{\varphi(\bar{\x})-\varphi(\bar{\x}+\y)}(1+e^{2\y^T\nabla\varphi(\bar{\x})})}{2} \\
   &\stackrel{(a)}\ge \frac{Z}{K}\frac{e^{-\y^T\nabla\varphi(\bar{\x})-\frac{L}{2}\norm{\y}^2}(1+e^{2\y^T\nabla\varphi(\bar{\x})})}{2} \\
   &= \frac{Z}{K}\frac{e^{-\frac{L}{2}\norm{\y}^2}(e^{\y^T\nabla\varphi(\bar{\x})}+e^{-\y^T\nabla\varphi(\bar{\x})})}{2}\\
   &\stackrel{(b)}\ge \frac{Z}{K}e^{-\frac{L}{2}\|\y\|^2}\\
   &\stackrel{(c)}\ge \frac{Z}{K}e^{-\frac{dL}{8}} > 0,\\\
\end{split}
\end{equation*}
where $(a)$ follows from (\ref{Lsmooth}) due to $L$-smoothness of $\varphi$, $(b)$ is due to AM-GM inequality and $(c)$ is because $\y\in [-\frac{1}{2},\frac{1}{2}]^d$.
\end{IEEEproof}

\begin{corollary}
If $\pi$ is a Gaussian density, then Algorithm~\ref{IMHR} produces a uniformly ergodic Markov chain.
\end{corollary}

\subsection{Effect of non-ideality of \textbf{ALG}}
 As defined in the previous section, \textbf{ALG} denotes the method used to draw samples from $\pi$ in Algorithm~\ref{IMHR}. For the analysis in this section, we suppose that \textbf{ALG} is an MCMC method. In practice, \textbf{ALG} could be methods like HMC or MALA. By non-ideality of \textbf{ALG}, we mean that the TVD between the probability distribution from which \textbf{ALG} generate samples and $\pi$ is nonzero. The non-ideality mentioned above can occur due to the finite time given for convergence in \textbf{ALG}. On account of this non-ideality, the proposed state in Algorithm~\ref{IMHR} will have a probability distribution different from the one used in the calculation of acceptance ratio (which is $\pi$). This alters the stationary distribution of Markov chain associated with Algorithm~\ref{IMHR}. 
 
 Suppose the Markov chain associated with \textbf{ALG} is geometrically ergodic for the stationary distribution $\pi$. In that case, we show in the following proposition that the error due to non-ideality of \textbf{ALG} can be bounded. For a discussion on the conditions under which methods like HMC and MALA are geometrically ergodic, we refer the reader to \cite{livingstone2016geometric},\cite{Langevin_roberts1996}.

\begin{prop}\label{inexactProp}
Let $\pi$ defined in (\ref{pi}) be such that $\varphi$ is L-smooth. Let $\B{t}\in\R^d$ be a fixed initial state of \textbf{ALG}. Suppose \textbf{ALG} satisfies the following geometric ergodicity condition.
\begin{equation}
    \|P^n(\B{t},\cdot)-\pi(\cdot)\|_{TV} \le V\rho^n,
\end{equation}
where $P$ is the transition kernel corresponding to \textbf{ALG}. ($V$ in the above expression may depend on the fixed initial state $\t$.) Also, let $P$ be such that $\pi\ll P(\t,\cdot)$ and $P(\t,\{\t\})>0$. Then for any $\x\in\R^d$, Algorithm~\ref{IMHR} generates a Markov chain with transition kernel $\bar{P}$ that satisfies the following inequality:
\begin{equation}\label{inexactBnd}
    \|\bar{P}^k(\x,\cdot)-\bar{\pi}(\cdot)\|_{TV} \le (1-C\delta)^k+\left(1+\frac{1}{C\delta}\right)\frac{V\rho^n}{\delta},
\end{equation}
where $\bar{\pi}$ is the target distribution defined in (\ref{piBar}), $\delta$ is obtained from (\ref{qbipi}), $n$ is the number of iterations of \textbf{ALG}, $k$ is the number of iterations of Algorithm~\ref{IMHR}, and $C$ is a constant that satisfies the following inequality:
\begin{equation}
    1-\frac{2V\rho^n}{\delta}\le C\le 1+\frac{2V\rho^n}{\delta}.
\end{equation}
\end{prop} 
\begin{IEEEproof}
Let us denote $P^n(\B{t},\cdot)$ by $q(\cdot)$. By geometric ergodicity of \textbf{ALG} we have,
\begin{equation}\label{geomErg_qpi}
    \|q-\pi\|_{TV} \le V\rho^n.
\end{equation}
Although \textbf{ALG} generates the proposed state from the distribution $q$, for calculating the acceptance ratio, we use the distribution $\pi$. Hence, the Markov chain generated by Algorithm~\ref{IMHR} has the following transition kernel
\begin{multline*}
    \bar{P}(\x,d\y) = q(d\y)\left(1\wedge\frac{\bar{\pi}(\y)\pi(\x)}{\bar{\pi}(\x)\pi(\y)}\right)\\ +\delta_{\x}(d\y)\left(1-\int_{\R^d}q(d\z)\left(1\wedge\frac{\bar{\pi}(\z)\pi(\x)}{\bar{\pi}(\x)\pi(\z)}\right)\right),
\end{multline*}
where $\delta_{\x}(\cdot)$ is the delta measure at $\x$. It is straightforward to verify using the detailed balance equation that the stationary distribution of the above Markov chain with transition kernel $\bar{P}$ is given by
\begin{equation}\label{statNu}
    \nu(d\x) = \frac{\bar{\pi}(\x)q(d\x)}{C\pi(\x)},
\end{equation}
where
\begin{equation}\label{NormC}
    C = \int_{\R^d} \frac{\bar{\pi}(\x)q(d\x)}{\pi(\x)}.
\end{equation}
Also, since $\varphi$ is $L$-smooth, from Proposition~\ref{uniErg}, we have
\begin{equation}\label{pipibar_prop}
    \frac{\pi(\x)}{\bar{\pi}(\x)}\ge \delta>0.
\end{equation}

The first step in this proof is to show that the above Markov chain with transition kernel $\bar{P}$ is uniformly ergodic. Then we show that its stationary distribution $\nu$ and probability density $\bar{\pi}$ are statistically close. These two results are combined to obtain the result stated in Proposition~\ref{inexactProp}.\\\\ 
\textit{Uniform Ergodicity of $\bar{P}$}:\\\\
From the expression for $\bar{P}$, for all $\x\in\R^d$ and all measurable $A\subseteq\R^d$, we have
\begin{equation*}
\begin{split}
    \bar{P}(\x,A) &\ge\int_A q(d\y)\left(1\wedge\frac{\bar{\pi}(\y)\pi(\x)}{\bar{\pi}(\x)\pi(\y)}\right)\\
    &\stackrel{(a)}= \int_A C\nu(d\y)\frac{\pi(\y)}{\bar{\pi}(\y)}\left(1\wedge\frac{\bar{\pi}(\y)\pi(\x)}{\bar{\pi}(\x)\pi(\y)}\right)\\
    &= \int_A C\nu(d\y)\left(\frac{\pi(\y)}{\bar{\pi}(\y)}\wedge\frac{\pi(\x)}{\bar{\pi}(\x)}\right)\\
    &\stackrel{(b)}\ge  C\delta\int_A\nu(d\y) = C\delta\nu(A),\\
\end{split}
\end{equation*}
where $(a)$ and $(b)$ are due to (\ref{statNu}) and (\ref{pipibar_prop}) respectively. Thus by Theorem~\ref{compState_uniform}, $\bar{P}$ is the transition kernel of a uniformly ergodic Markov chain. Therefore,
\begin{equation}
    \|\bar{P}^k(\B{t},\cdot)-\nu(\cdot)\|_{TV} \le (1-C\delta)^k.
\end{equation}
Next, we find a bound on the value of $C$. Note that from the assumption $P(\t,\{\t\})>0$, it follows that for any measurable set $A\subseteq\R^d$ and integer $n$, $P^n(\t,A)>0$ whenever $P(\t,A)>0$. Therefore, $P(\t,\cdot)\ll q$. This, together with the assumption $\pi\ll P(\t,\cdot)$, allows us to conclude that $\pi\ll q$. Therefore, we have the following:
\begin{equation}\label{splitOne}
\begin{split}
    1 = \int_{\R^d}\bar{\pi}(d\x) &\stackrel{(a)}= \int_{\R^d}\frac{\bar{\pi}(\x)}{\pi(\x)}\pi(d\x)\\
    &\stackrel{(b)}= \int_{\R^d}\frac{\bar{\pi}(\x)}{\pi(\x)}\frac{d\pi}{dq}(\x)q(d\x),
\end{split}
\end{equation}
where $(a)$ follows from the fact that $\pi$ and $\bar{\pi}$ are density functions which are positive everywhere and $(b)$ is due to the absolute continuity of $\pi$ with respect to $q$. Then,
\begin{equation*}
    \begin{split}
        |C-1| &\stackrel{(a)}= \left|\int_{\R^d} \frac{\bar{\pi}(\x)}{\pi(\x)}q(d\x)-\int_{\R^d}\frac{\bar{\pi}(\x)}{\pi(\x)}\frac{d\pi}{dq}(\x)q(d\x)\right|\\
        &= \left|\int_{\R^d} \frac{\bar{\pi}(\x)}{\pi(\x)}\left(1-\frac{d\pi}{dq}(\x)\right)q(d\x)\right|\\
        &\le \int_{\R^d}\frac{\bar{\pi}(\x)}{\pi(\x)}\left|1-\frac{d\pi}{dq}(\x)\right|q(d\x)\\
        &\le \frac{1}{\delta}\int_{\R^d}\left|1-\frac{d\pi}{dq}(\x)\right|q(d\x)\\
        &\stackrel{(b)}= \frac{2}{\delta}\norm{q-\pi}_{TV}\\
        &\le \frac{2V\rho^n}{\delta},
    \end{split}
\end{equation*}
where $(a)$ follows from (\ref{NormC}) and (\ref{splitOne}), and $(b)$ is due to the alternate definition of TVD given in (\ref{TVD_radonNik}). Therefore, we have
\begin{equation}\label{Cbound}
    \begin{split}
    \delta-2V\rho^n &\le C\delta \le \delta+2V\rho^n\\
    \implies 1-\frac{2V\rho^n}{\delta} &\le C\le 1+\frac{2V\rho^n}{\delta}.
    \end{split}
\end{equation}\\
\textit{TVD between $\nu$ and $\bar{\pi}$}:\\\\
 Now we will show that $\nu$ and $\bar{\pi}$ are statistically close probability distributions.
\begin{equation}
    \begin{split}
        \|\nu-\bar{\pi}\|_{TV} &\stackrel{(a)}= \frac{1}{2}\int_{\R^d} \left|\frac{d\nu}{dq}(\x)-\frac{d\bar{\pi}}{dq}(\x)\right|q(d\x)\\
        &= \frac{1}{2}\int_{\R^d} \left|\frac{\bar{\pi}(\x)}{C\pi(\x)}-\frac{\bar{\pi}(\x)}{\pi(\x)}\frac{d\pi}{dq}(\x)\right|q(d\x)\\
        &= \frac{1}{2C}\int_{\R^d}\frac{\bar{\pi}(\x)}{\pi(\x)} \left|1-C\frac{d\pi}{dq}(\x)\right|q(d\x)\\
        &\le \frac{1}{2C\delta}\int_{\R^d} \left|1-C\frac{d\pi}{dq}(\x)\right|q(d\x)\\
        &\le \frac{1}{2C\delta}\int_{\R^d} \left(C\left|1-\frac{d\pi}{dq}(\x)\right|+\left|C-1\right|\right)q(d\x)\\
        &= \frac{1}{\delta}\|q-\pi\|_{TV} + \frac{|C-1|}{2C\delta}\\
        &\stackrel{(b)}\le \frac{V\rho^n}{\delta}+\frac{V\rho^n}{C\delta^2},
    \end{split}
\end{equation}
where $(a)$ is due to the alternate definition of TVD given in (\ref{TVD_radonNik}), and $(b)$ is due to (\ref{geomErg_qpi}) and (\ref{Cbound}).\\

Finally using triangle inequality, we have
\begin{equation}
    \begin{split}
        \|\bar{P}^k(\B{t},\cdot)-\bar{\pi}(\cdot)\|_{TV} &\le \|\bar{P}^k(\B{t},\cdot)-\nu(\cdot)\|_{TV}+\|\nu-\bar{\pi}\|_{TV}\\
        &\le (1-C\delta)^k+(1+\frac{1}{C\delta})\frac{V\rho^n}{\delta}.
    \end{split}
\end{equation}
\end{IEEEproof}

\section{Simulation Results}\label{SimRes}
This section illustrates the speed of convergence of Algorithm~\ref{IMHR} to $P_{\Z^d}$. For this, ideally we would like to show plots of TVD as a function of the number of iterations.
However, evaluating distance between high dimensional probability distributions is computationally hard. So, in our simulations, we compute an entity $\text{TVD}_\text{m}$ instead of TVD. We compute $\text{TVD}_\text{m}$ as follows: Initialize Algorithm~\ref{IMHR} with a fixed point in the state space. Then we run $t$ iterations of Algorithm~\ref{IMHR}. Repeat this 100,000 times for each value of $t$. For each $t$, use these samples to form $d$ 1-dimensional histograms, one for each coordinate. We call the distributions obtained by normalizing the histograms as the empirical marginal distributions, denoted by $h^i(\z)$ for $1\le i\le d$ and $\z\in\Z$. We denote the $i^{th}$ marginal distribution of $P_{\Z^d}$ by $P^i_{\Z}$. If $P^i_{\Z}$ is not available in closed form, we estimate it using an MCMC method (see Appendix~\ref{distEst} for the exact algorithm used), with sufficient time given for convergence. Calculate the Total Variation Distance between $h^i$ and $P^i_{\Z}$ using the following formula:
\begin{equation*}
    \text{TVD}(i) = \frac{1}{2}\sum_{\z\in\Z}|h^i(\z)-P^i_{\Z}(\z)| \quad \text{for } 1\le i\le d.
\end{equation*}
Finally, $\text{TVD}_\text{m}$ is the maximum of TVD's calculated for marginal distributions.
\begin{equation*}
    \text{TVD}_\text{m} = \max\{\text{TVD}(i): 1\le i\le d\}.
\end{equation*}
We plot $\text{TVD}_\text{m}$ for different values of $t$. The $\text{TVD}_\text{m}$ vs. $t$ plots depicts the number of iterations required for Algorithm~\ref{IMHR} to converge to its stationary distribution.

In another simulation, we plot the autocorrelation function of the time series $\x_0,\x_1,\cdots,\x_N$ obtained using Algorithm~\ref{IMHR}. In many instances, autocorrelation plots have been used to assess the number of iterations of the Markov chain required to produce two almost independent samples\cite{handbook_MCMC}. The definition of the autocorrelation function that we use is as follows:
\begin{equation*}
        \text{ACF}(\tau) = \frac{\sum_{t=1}^{N-\tau}\x_t^T\x_{t+\tau}}{\sum_{t=1}^{N}\x_t^T\x_{t}},
\end{equation*}
where $\x_t$ is the state of the Markov chain at iteration $t$ and $N$ is the total number of samples in the time series. Now we present results of these simulations for different probability densities $\pi$.

\subsubsection{Isotropic Gaussian distribution}
\begin{figure}[htb]
\centering
\includegraphics[scale=0.47]{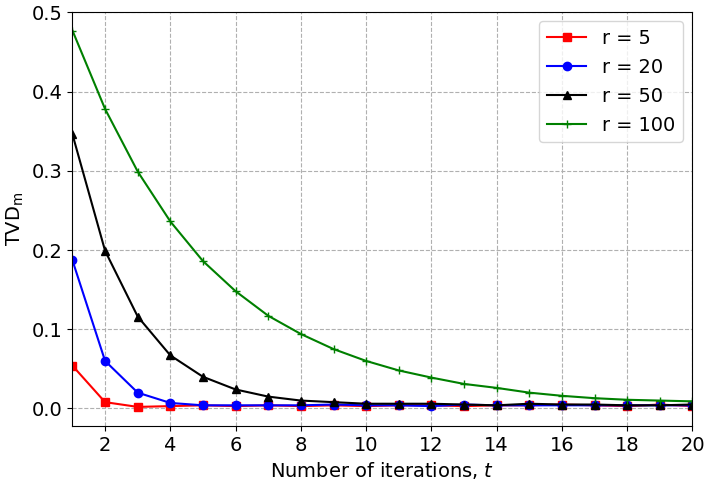}
\caption{$\text{TVD}_\text{m}$ vs. Number of Iterations ($t$) for isotropic Gaussian}
\label{tvd_iso}
\end{figure}

\begin{figure}[htb]
\centering
\includegraphics[scale=0.47]{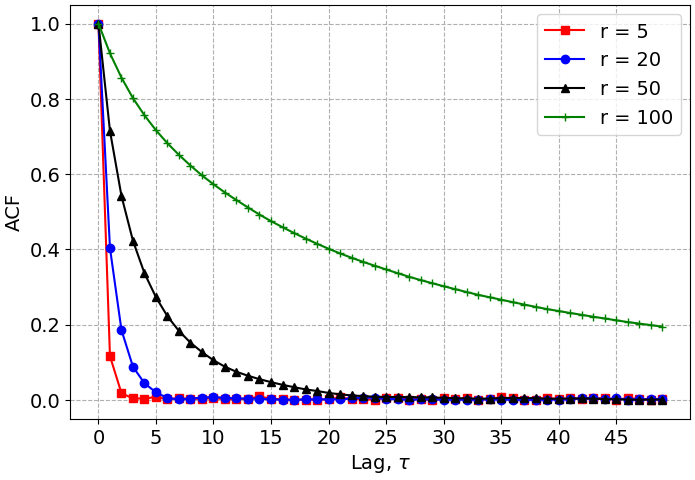}
\caption{ACF vs $\tau$ for isotropic Gaussian}
\label{tmix_iso}
\end{figure}

We first consider the case when $\pi$ is an isotropic Gaussian density. The potential function $\varphi$ of an isotropic Gaussian density is $\frac{1}{\sigma^2}$-smooth, where $\sigma^2$ is the variance of the Gaussian. Therefore, from Proposition~\ref{uniErg}, we see that the factor that governs the rate of convergence of Algorithm~\ref{IMHR} is $r = \frac{d}{\sigma^2}$. In this simulation, we fix $\sigma^2=1$ and vary dimension $d$ to get different values of $r$. We use state $\B{0}$ as the initial state of the algorithm. $\text{TVD}_\text{m}$ vs. $t$ for different values of $r$ is shown in Figure~\ref{tvd_iso}. Autocorrelation vs. $\tau$ is shown in Figure~\ref{tmix_iso}. The number of samples ($N$) used to calculate the autocorrelation function is 10,000.

\subsubsection{Gaussian distribution on the Leech lattice}
\begin{figure}[htb]
\centering
\includegraphics[scale=0.47]{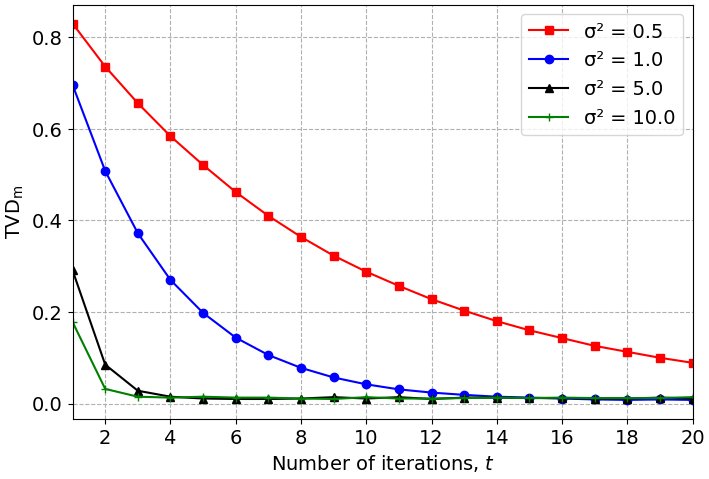}
\caption{$\text{TVD}_\text{m}$ vs. Number of Iterations ($t$) for a Gaussian distribution on the Leech lattice}
\label{tvd_leech}
\end{figure}

Next, we consider a lattice Gaussian distribution supported on the Leech lattice with dimension equal to 24 (see (\ref{genMat}) for the generator matrix of the Leech lattice). This simulation illustrates the performance of Algorithm \ref{IMHR} for non-isotropic Gaussian. The Leech lattice induces a highly skewed lattice Gaussian distribution on $\Z^d$. The density $\pi$ now takes the following form:
\begin{equation*}
    \pi(\x) = Me^{-\frac{\|\B{B}\x\|^2}{2\sigma^2}} \quad \text{for all } \x\in\R^d,
\end{equation*}
where $\B{B}$ is the generator matrix of the Leech lattice and $M$ is the normalization constant. We plot $\text{TVD}_\text{m}$ vs. $t$ for different values of $\sigma^2$. This is shown in the Figure~\ref{tvd_leech}. State $\B{0}$ was used as the initial state of the algorithm.

\subsubsection{Perfect Security distribution}\label{PSD}
\begin{figure}[htb]
\centering
\includegraphics[scale=0.47]{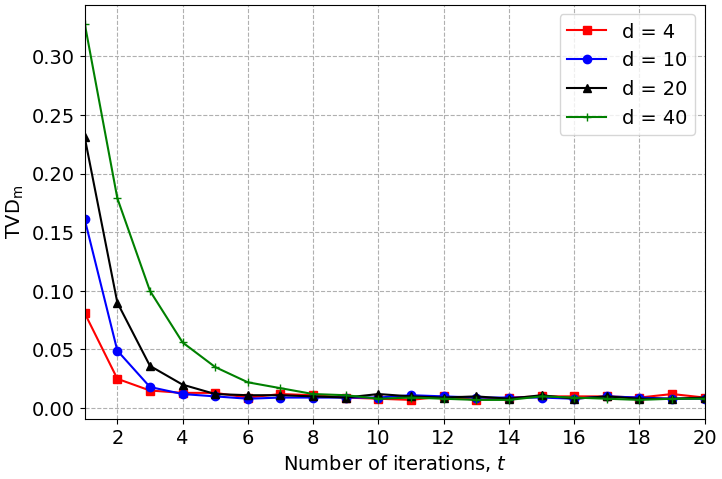}
\caption{$\text{TVD}_\text{m}$ vs. Number of Iterations ($t$) for Perfect Security distribution}
\label{tvd_minv}
\end{figure}

Finally, we present the simulation results when $\pi$ is the following probability density which was used to achieve \textit{perfect security} in \cite{Vatedka}. The probability density function of a ``Perfect Security distribution'' is given by: 
\begin{equation*}
    \pi(\x) = M\left(\frac{\Omega_d(\frac{\norm{\x}}{2\rho})}{j_{\frac{d-2}{2}}^2-\frac{\norm{\x}^2}{4\rho^2}}\right)^2 \quad \text{for all } \x\in\R^d,
\end{equation*}
where
\begin{equation*}
    \Omega_d(u) = \left(\frac{2}{u}\right)^{\frac{d-2}{2}}J_{\frac{d-2}{2}}(u),
\end{equation*}
$J_k(u)$ is the Bessel function of $k^{th}$ order and $j_k$ is the first zero of $k^{th}$ order Bessel function and $M$ is the normalization constant. Let $\X=[X_1,X_2,\cdots,X_d]$ be a random vector with perfect security probability distribution. Then, the variance of each component $X_i$ is given by the following equation\cite{Ehm_convRoots}:
\begin{equation*}
    \text{Var}(X_i) = \frac{4\rho^2j_{\frac{d-2}{2}}^2}{d}.
\end{equation*}
We use HMC (see Appendix~\ref{HMC} for parameters used) to sample from this continuous density $\pi$.  We plot $\text{TVD}_\text{m}$ vs. $t$ for different values of $d$. We fix the value of $\rho$ such that the variance of the distribution is $1$ for each value of $d$. The $\text{TVD}_\text{m}$ vs. $t$ plot is shown in Figure~\ref{tvd_minv}. State $\B{0}$ was used as the initial state of the algorithm.
\section{Discussion}\label{discussion}
We presented a simple MCMC algorithm to draw samples from lattice distributions. As demonstrated through the Perfect Security distribution sampling, Algorithm~\ref{IMHR} can sample from distributions beyond lattice Gaussians. To the best of our knowledge, prior to this work, there were no efficient algorithms known to generate samples from lattice distributions other than lattice Gaussians. The main feature of Algorithm~\ref{IMHR}, which makes it competitive even among the lattice Gaussian sampling algorithms, is its computational efficiency.  The computations in Algorithm~\ref{IMHR} are vector operations, which is highly optimized when current linear algebra libraries (for instance, OpenBLAS or Intel MKL) are used for implementation. Most of the algorithms currently available for sampling from a lattice Gaussian do coordinate-wise sequential sampling using 1-dimensional lattice Gaussian samplers. This method is inefficient when the lattice dimension under consideration is large.

\begin{figure}[htb]
\centering
\includegraphics[scale=0.5]{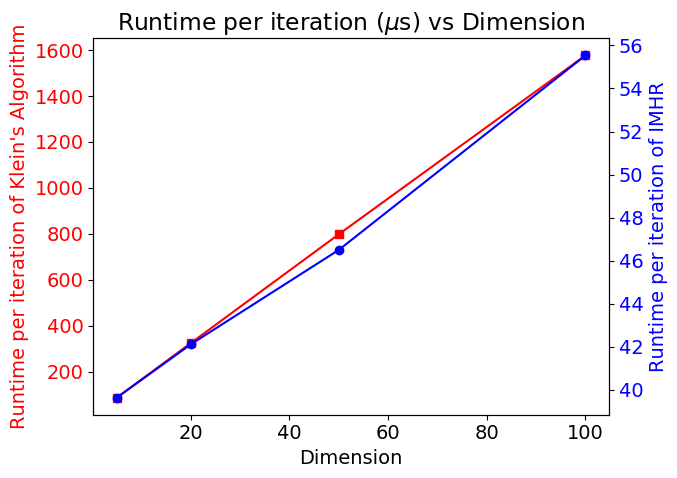}
\caption{Run-time vs Dimension for isotropic lattice Gaussian}
\label{runtime}
\end{figure}

A popular algorithm for sampling from lattice Gaussians is Klein's algorithm\cite{klein},\cite{Gentry_Peikert}. In Figure~\ref{runtime}, we compare the run-time per iteration of Klein's algorithm with Algorithm~\ref{IMHR} for different values of dimension when the desired distribution is a lattice Gaussian on $\Z^d$ with variance parameter $\sigma = 1$. This experiment was run in a python environment on a machine with Intel i7-6700 @ 3.40GHz CPU and 8GB RAM. It is clear from Figure~\ref{runtime} that the scaling of the run-time per iteration with dimension is much better for Algorithm~\ref{IMHR}. However, multiple iterations of Algorithm~\ref{IMHR} are required to generate a sample approximately from the stationary distribution. From Figure~\ref{tvd_iso}, we can obtain the minimum number of iterations of Algorithm~\ref{IMHR} required to bring the $\text{TVD}_\text{m}$ below a small number. Multiplying the run-time per iteration of Algorithm~\ref{IMHR} by the minimum number of iterations, we see that the run-time required to generate a sample from lattice Gaussian is comparable for Klein's algorithm and Algorithm~\ref{IMHR}. For example, Algorithm~\ref{IMHR} takes 13 iterations to bring the $\text{TVD}_\text{m}$ below 0.005 when dimension equals 50. Run-time per iteration for Algorithm~\ref{IMHR} is 46  $\mu s$ when dimension equals 50. This implies a total run-time of 598 $\mu s$ to generate a sample approximately from the stationary distribution. Klein's algorithm requires just one iteration to generate a sample from a lattice Gaussian distribution. However, it takes 798 $\mu s$ per iteration.

\begin{figure}[htb]
\centering
\includegraphics[scale=0.5]{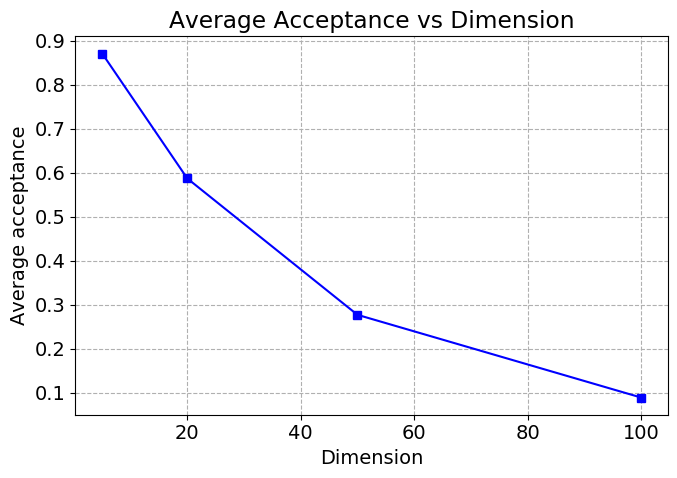}
\caption{Average Acceptance vs Dimension for isotropic lattice Gaussian}
\label{av_acc}
\end{figure}

From Proposition~\ref{uniErg}, we see that when $\pi$ is an isotropic Gaussian density with variance equal to $\sigma^2$, the TVD between the probability distribution after $k^{th}$ iteration of Algorithm~\ref{IMHR} and the stationary distribution is upper bounded by $(1-\frac{Z}{K}e^{-\frac{d}{8\sigma^2}})^k$. This indicates that our algorithm may not be well suited for distributions with very low variance and high dimension. Figures \ref{tvd_iso} and \ref{tvd_leech} validate this by illustrating that convergence is slow for low variance and high dimension cases. In simulations, we observe that at high dimensions, the average acceptance, which is the fraction of iterations in which the proposed state is accepted, becomes very low for Algorithm~\ref{IMHR}.  Figure~\ref{av_acc} shows the degradation of average acceptance with dimension. A low acceptance ratio makes the Independent Metropolis-Hastings algorithm inefficient due to frequent rejection of the proposed state. Therefore, at very high dimensions, we suggest using the Metropolis-within-Gibbs strategy\cite{tierney1994}. In the Metropolis-within-Gibbs algorithm, the number of variables updated at a time, determines the average acceptance.

\bibliographystyle{IEEEtran}
\bibliography{IEEEabrv,refs}
\newpage
\appendices
\section{Piece-wise Constant approximation for Gaussian density}\label{PWConstApprox}
In this section, we substantiate the claim in Section \ref{IMHR_section} that the  choice of $\bar{\pi}(\x) = \pi([\x])$ can give rise to a Markov chain which is not uniformly ergodic.  We show this for a simple case where $\pi$ is a 1-dimensional Gaussian density. From Theorem 2.1 in \cite{mengersen1996}, it follows that, if $\text{ess}\inf \frac{\pi(\x)}{\bar{\pi}(\x)} = 0$ with respect to $\bar{\pi}$ measure, then Independent Metropolis Hastings algorithm is not even geometrically ergodic. Let $\pi$ be a  1-dimensional Gaussian density and $\bar{\pi}(x) = \pi([x])$. Let $\bar{x}$ denote $[x]$ and let $y = x-\bar{x}$. Then,
\begin{equation}\label{essInf}
\begin{split}
    \frac{\pi(x)}{\Bar{\pi}(x)} &= M\frac{e^{-(\bar{x}+y)^2}}{e^{-\bar{x}^2}}\\
    &= Me^{-2\bar{x}y}e^{-y^2},
\end{split}
\end{equation}
where $M$ is a constant. By definition, essential infimum of $\frac{\pi(x)}{\Bar{\pi}(x)}$ with respect to $\bar{\pi}$ measure is the greatest number $a$ such that the set,
\begin{equation*}
    A = \{x\in\R:\frac{\pi(x)}{\Bar{\pi}(x)}<a\}
\end{equation*}
has zero $\bar{\pi}$-measure. It is clear from (\ref{essInf}) that, by choosing a large value for $x$, $\frac{\pi(x)}{\Bar{\pi}(x)}$ can be made arbitrary close to 0 within a set of nonzero $\bar{\pi}$ measure.
\begin{equation*}
    \implies \text{ess}\inf {\frac{\pi(x)}{\Bar{\pi}(x)} = 0 }.
\end{equation*}
This shows that for $\bar{\pi}(\x) = \pi([\x])$, Independent Metropolis Hastings algorithm need not even be geometrically ergodic.

\section{Other MCMC methods used}\label{othMethods}
\subsection{MCMC method used in the estimation of marginal distributions}\label{distEst}
This section elaborates on the MCMC method used for estimating marginal distributions $P^i_{\Z}$ required to calculate $\text{TVD}_\text{m}$ in Section \ref{SimRes}. We use the Random Walk Metropolis (RWM) algorithm to estimate the marginal distributions. This is a Metropolis-Hastings algorithm in which proposal density $q(\x,\y)$ is a function of $\norm{\y-\x}$. We refer an interested reader to \cite{JARNER2000341},\cite{roberts1997} for more on random walk Metropolis algorithms. Target distribution $\bar{\pi}$ used in this algorithm is the following piece-wise constant density derived from $\pi$.
\begin{equation*}
    \bar{\pi}(\x) = \pi([\x])\quad \text{for all } \x\in\R^d.
\end{equation*}
Due to the symmetric nature of the proposal density, the acceptance ratio takes the following simple form:
\begin{equation*}
    \alpha(\x,\y) = 1\wedge\frac{\bar{\pi}(\y)}{\bar{\pi}(\x)}.
\end{equation*}
In particular, we use the random walk Metropolis algorithm with proposal density $q(\x,\cdot)$ being a Gaussian density with mean $\x$ and covariance matrix $\Sigma$. We choose $\Sigma$ to be proportional to the covariance matrix of $\pi$. 

 Algorithm \ref{RWM} describes the steps involved in the random walk Metropolis. We do 500 iterations of this algorithm to give it enough time to converge to the stationary distribution and thereby generate one sample. We use 200,000 such samples to form the histogram for each co-ordinate, which gives us the estimate of marginal distributions.

\begin{algorithm}\label{RWM}
\SetAlgoLined
\KwIn{ $\pi,\Sigma,\X_0$}
\KwOut{Sample from a distribution statistically close to $P_{\Z^d}$}
 \For{$t=1,2,\ldots$}{
  Let $\x$ denote the state of $\X_{t-1}$\;
  Generate $\w$ from $\mathcal{N}(0,\Sigma)$\;
  $\y\leftarrow\x+\w$\;
  Round $\y$ to its nearest point in $\Z^d$ to get $\bar{\y}$\;
  Round $\x$ to its nearest point in $\Z^d$ to get $\bar{\x}$\;

  Calculate acceptance ratio $\alpha(\x,\y) = 1\wedge\frac{\pi(\bar{\y})}{\pi(\bar{\x})}$\;
  Generate a sample $u$ from $U[0,1]$\;
  \eIf{$u\le \alpha(\x,\y)$}{
   let $\X_t=\y$\;
   }{
   $\X_t=\x$\;
  }
  \If{$t>t_{\text{mix}}(\epsilon;\X_0)$}{
  Round $\X_t$ to its nearest point in $\Z^d$ to get $\bar{\X}_t$\;
  }
 }
 \caption{Random Walk Metropolis Algorithm}
\end{algorithm}

\subsection{Hamiltonian Monte Carlo (HMC)}\label{HMC}
HMC is used to generate samples from perfect security distribution in section \ref{PSD}. We refer the reader to \cite{neal2012mcmc} for an exposition on HMC. The input parameters to HMC are the number of Leapfrog steps ($L$) and the Leapfrog step-size ($\epsilon$). The values of $L$ and $\epsilon$ used in our simulation are as given below:
\begin{align*}
    L &= \left\lfloor5\left(\frac{2}{d}\right)^\frac{1}{4}\right\rfloor,\\
    \epsilon &= 1.2\left(\frac{2}{d}\right)^\frac{1}{4}.
\end{align*}
Inside HMC, we resample momentum variables from an isotropic Gaussian density with variance equal to 9. We use five iterations of HMC to approximately generate a sample from the perfect security distribution.

\setcounter{MaxMatrixCols}{24}
\begin{figure*}[htb]
\begin{equation}\label{genMat}
B = \frac{1}{\sqrt{8}}\begin{bmatrix}
8&4&4&4&4&4&4&2&4&4&4&2&4&2&2&2&4&2&2&2&0&0&0&-3\\
0&4&0&0&0&0&0&2&0&0&0&2&0&2&0&0&0&0&0&2&2&0&0&1\\
0&0&4&0&0&0&0&2&0&0&0&2&0&0&2&0&0&2&0&0&2&0&0&1\\
0&0&0&4&0&0&0&2&0&0&0&2&0&0&0&2&0&0&2&0&2&0&0&1\\
0&0&0&0&4&0&0&2&0&0&0&0&0&2&2&2&0&2&2&2&2&0&0&1\\
0&0&0&0&0&4&0&2&0&0&0&0&0&2&0&0&0&0&2&0&0&0&0&1\\
0&0&0&0&0&0&4&2&0&0&0&0&0&0&2&0&0&0&0&2&0&0&0&1\\
0&0&0&0&0&0&0&2&0&0&0&0&0&0&0&2&0&2&0&0&0&0&0&1\\
0&0&0&0&0&0&0&0&4&0&0&2&0&2&2&2&0&2&2&2&2&2&2&1\\
0&0&0&0&0&0&0&0&0&4&0&2&0&2&0&0&0&2&0&0&0&2&0&1\\
0&0&0&0&0&0&0&0&0&0&4&2&0&0&2&0&0&0&2&0&0&0&2&1\\
0&0&0&0&0&0&0&0&0&0&0&2&0&0&0&2&0&0&0&2&0&0&0&1\\
0&0&0&0&0&0&0&0&0&0&0&0&4&2&2&2&0&0&0&0&2&2&2&1\\
0&0&0&0&0&0&0&0&0&0&0&0&0&2&0&0&0&0&0&0&0&2&0&1\\
0&0&0&0&0&0&0&0&0&0&0&0&0&0&2&0&0&0&0&0&0&0&2&1\\
0&0&0&0&0&0&0&0&0&0&0&0&0&0&0&2&0&0&0&0&0&0&0&1\\
0&0&0&0&0&0&0&0&0&0&0&0&0&0&0&0&4&2&2&2&2&2&2&1\\
0&0&0&0&0&0&0&0&0&0&0&0&0&0&0&0&0&2&0&0&0&2&0&1\\
0&0&0&0&0&0&0&0&0&0&0&0&0&0&0&0&0&0&2&0&0&0&2&1\\
0&0&0&0&0&0&0&0&0&0&0&0&0&0&0&0&0&0&0&2&0&0&0&1\\
0&0&0&0&0&0&0&0&0&0&0&0&0&0&0&0&0&0&0&0&2&2&2&1\\
0&0&0&0&0&0&0&0&0&0&0&0&0&0&0&0&0&0&0&0&0&2&0&1\\
0&0&0&0&0&0&0&0&0&0&0&0&0&0&0&0&0&0&0&0&0&0&2&1\\
0&0&0&0&0&0&0&0&0&0&0&0&0&0&0&0&0&0&0&0&0&0&0&1\\
\end{bmatrix}
\end{equation}
\end{figure*}

\end{document}